# Bismuth and Antimony Based Oxyhalides and Chalcohalides as Potential Optoelectronic Materials

Zhao Ran[1,†], Xinjiang Wang[1,†], Yuwei Li[2], Dongwen Yang[1], Xin-Gang Zhao[1], Koushik Biswas[3], David J. Singh[2,*], and Lijun Zhang[1,*]

[1]*Key Laboratory of Automobile Materials of MOE, State Key Laboratory of Superhard Materials, and College of Materials Science, Jilin University, Changchun 130012, China*
[2]*Department of Physics and Astronomy, University of Missouri, Columbia, MO 65211-7010 USA*

[3]*Department of Chemistry and Physics, Arkansas State University, State University, AR 72467, USA*

[†]These authors contributed equally.

[*]Address correspondence to: singhdj@missouri.edu or lijun_zhang@jlu.edu.cn

## ABSTRACT

In the last decade the n*s*$^2$ cations (*e.g.*, $Pb^{2+}$ and $Sn^{2+}$) based halides have emerged as one of the most exciting new classes of optoelectronic materials, as exemplified by for instance hybrid perovskite solar absorbers. These materials not only exhibit unprecedented performance in some cases, but they also appear to break new ground with their unexpected properties, such as extreme tolerance to defects. However, because of the relatively recent emergence of this class of materials, there remain many yet to be fully explored compounds. Here we assess a series of bismuth/antimony oxyhalides and chalcohalides using consistent first principles methods to ascertain their properties and obtain trends. Based on these calculations, we identify a subset consisting of three types of compounds that may be promising as solar absorbers, transparent conductors, and radiation detectors. Their electronic structure, connection to the crystal geometry, and impact on band-edge dispersion and carrier effective mass are discussed.

## INTRODUCTION

Halides containing high-*Z* n*s*$^2$ cations (with electron configurations of $[d^{10}]s^2p^0$, *e.g.*, $Tl^+$, $Pb^{2+}$ and $Bi^{3+}$) have been found to possess unexpectedly good carrier transport properties. Depending on specific optical band gaps and other features of the band structure, they can be useful optoelectronic materials for various applications such as radiation detection, solar cells, etc.[1-6] For instance, $PbI_2$ and some Tl-based halides have shown excellent performance as semiconductor radiation detectors, in which carriers generated by γ- or X-ray radiation can be efficiently collected over millimeter or even longer distances.[1–4] Meanwhile solar cells based on Pb-based organic-inorganic hybrid halide perovskites have achieved power conversion efficiency well above 20%.[5,6] The prototypical material, $CH_3NH_3PbI_3$ is exceptional in having an extremely large dielectric constant on the border of ferroelectricity, suitable band gap for solar absorber application and low hole and electron effective masses.[7-14] The physical mechanisms underlying these optoelectronic halide materials, are not yet fully established, but can at least in part be attributed to the unique chemistry of the n*s*$^2$ lone-pair state. Factors that have been discussed include formation of defects with lower charge states, high dielectric constants, and shallow defect levels among others.[7-15] Assessment can be made experimentally using time resolved experiments for carrier lifetimes.[14]

In these compounds the cation-*s* state moves down in energy (forming the lone-pair state), separating from the cation-*p* states (due to the Mass-Darwin effect), so that there is substantial charge transfer only from the cation-*p* to anion-*p* states, rather than the cation-*s* to anion-*p* states. Thus, the valence bands are mainly formed by anti-bonding combinations of cation-*s* and anion-*p* states; the conduction bands are dominated by the spatially extended cation-*p* states, hybridizing also in the anti-bonding pattern with the anion-*p* states. This leads to simultaneously dispersive valence and conduction bands (favorable to high mobility), and defect tolerance. Furthermore, the resulting significant cross-band-gap hybridization gives rise to enhanced Born effective charges, analogous to classical ferroelectric materials, such as $BaTiO_3$. This enhances the dielectric constant, which effectively screens defects and is beneficial for carrier transport.[15,16]

In addition to the n$s^2$ cations containing halides, the corresponding oxyhalides and chalcohalides have been also considered as optoelectronic materials. In 2012, experimental studies by Hahn *et al.* found BiSI and BiSeI to be *n*-type, with high absorption coefficients.[17,18] However, devices based on BiSI have up to now shown poor performances.[18] Ganose *et al.*, based on density functional calculations, attributed the poor performance in the reported BiSI-based devices to band misalignments and suggested alternative device architectures.[19] Theoretical studies of Bi(III) chalco-halides have suggested *n*-BiSeBr, *p*-BiSI, and *p*-BiSeI as photovoltaic materials, BiSeBr and BiSI as room-temperature radiation detection materials, and BiOBr as a *p*-type transparent conducting material.[10]

The purpose of this paper is to explore these compounds more broadly seeking identification of new optoelectronic compounds and trends. Specifically, we assess a series of bismuth/antimony oxyhalides and chalcohalides. It is important to note the rich diversity of chemical compositions and geometrical structures in this class of compounds, which supports the notion that interesting properties are yet to be discovered in them. The compound-intrinsic properties (e.g. thermodynamic stability, electronic structure, effective masses of electron/hole and optical absorption) of 36 relatively simple compounds obtained from the Inorganic Crystal Structure Database (ICSD) have been systematically investigated through density functional calculations. Based on the results, we found a series of candidates suitable for optoelectronic applications, including two types of compounds (BiSeCl/Br/I and $Bi_3Se_4Br$) as potential solar absorber materials, three types of compounds ($Bi_3Se_4Br$, BiSeCl/Br/I and BiOI) as potential radiation detection materials, and three types of compounds (BiOCl/Br, SbOF, and $Sb_4O_5Cl_2/Br_2$) as potential *p*-type transparent conducting materials.

## RESULTS

### Properties of the 36 Bi/Sb oxyhalides and chalcohalides

The calculated HSE+SOC band gaps of representative compounds are listed in Table I, compared with available experimental results. One sees that the calculated gap values are in very good agreement with corresponding experimental data for most of the compounds except the tellurides, where the band gaps are overestimated compared to experiment with the HSE functional. In the following, we only consider the oxyhalides and chalcohalides not containing Te. It may also be noted that besides increasing the band gap relative to PBE+SOC, HSE+SOC calculations may lead to different band dispersions, for example changes in the band curvatures and effective masses. We calculated HSE+SOC band structures to compare with PBE+SOC for some compounds, specifically the compounds that we identify as potential solar absorbers. As seen (Supplementary Fig. S1), while there are small differences, these changes are minor for the compounds studied.

The calculated formation energies, band gaps and effective masses of electrons/holes for the 36 Bi and Sb based compounds are given in Fig. 1 and Supplementary Table S1. Here, the formation energy is defined as the energy difference between a compound and its constituent elemental solids. It is seen that all the compounds possess negative formation energies (upper panel in Fig. 1), indicating that they are stable against decomposition to constituent elemental solids, as expected, considering that they are experimentally known compounds. Generally, the stabilities of the compounds increase as the halogens change from I to F or the chalcogens change form Te to O. This is the trend of increasing anion electronegativity, and therefore increasing electronegativity difference between anions and cations, which normally leads to increasing stability for ionic compounds. We also calculated the thermodynamic stability of the first category of compounds (BiSe/Cl/Br/I and $Bi_2Se_4Br$) with respect to competing phases, including structural optimization with inclusion of the van der Waals interaction based on the optB86 functional[35] (important for the structure of the binaries) and find marginal stability against decomposition into binaries (Supplementary Table S2).

As shown in the middle panel of Fig. 1, the calculated band gaps of these Bi and Sb based compounds cover a wide range from ~1 to ~5 eV. There is a general trend of increasing band gap with increasing electronegativity difference between constituent cations and anions, as indicated in Fig. 2(a). There is also a clear trend in the bond valence sum (BVS)[36] plot shown in Fig. 2(c), where the band gaps increase as the BVS values of Bi or Sb approach their the nominal valence of three, reflecting more ideal bonding geometries.

The effective masses of electrons and holes of these compounds are shown in the lower panel of Fig. 1. In most cases the electron effective masses are lower than those of holes, but there is no obvious relationship between the effective masses, and the electronegativity difference [Fig. 2(b)] or the cation BVS [Fig. 2(d)]. Strikingly, many of these compounds possess electron or hole effective masses lower than or around the rest mass of electrons ($m_0$), indicating the possibility of good carrier mobility. The large span of the band gaps along with low electron or hole effective masses suggests opportunities for various optoelectronic applications based on compounds in this class.

The calculated dielectric constants, as obtained from PBE calculations, are also large for almost all of the compounds as shown in Fig. 3 (with the explicit numbers in Supplementary Table S3). Note that the main contributions to the dielectric constants are from the lattice, reflecting the mechanism discussed above. These high dielectric constants are then expected to lead to screening of defect potentials and suppression of point defect scattering.

Suitable optical band gap is a primary prerequisite for specific optoelectronic applications, e.g. gaps in range 1.0-1.7 eV are optimum for solar absorber materials. A larger gap, greater than 1.5 eV is essential for room-temperature radiation detection applications in order to minimize noise due to thermally generated carriers. Band gaps near to or larger than 3 eV are needed for transparent conducting materials. In addition, small effective mass of electrons or holes is also necessary for good carrier mobility. These two compound properties enable us to identify candidates as potential optoelectronic materials, which then deserve further theoretical or experimental studies concerning their dopability, dielectric and other properties.

**Potential solar absorbers**

Four out of 36 compounds (BiSeCl, BiSeBr, BiSeI and $Bi_3Se_4Br$) are found to possess optical band gaps in range of ~1.0-1.7 eV and electron or hole effective masses lower than $m_0$. The band gaps and average carrier effective masses of the four compounds are given in histogram in Fig. 4(a). There are two distinct crystal structures in this set of four compounds: BiSeCl, BiSeBr and BiSeI possess a crystal structure of space group *Pnma* (No. 62), and $Bi_3Se_4Br$ with space group *C*2/*m* (No. 12).

The iso-structural compounds, BiSeCl, BiSeBr and BiSeI have similar electronic properties, including similar optical band gaps of approximately 1.47-1.48 eV, and similar transport effective masses of ~0.53-0.57 $m_0$ and ~2.19-2.98 $m_0$, for electrons and holes, respectively. This reflects the dominant role of the chalcogen in the upper valence band electronic structures. For comparison the cubic phase of $CH_3NH_3PbI_3$ has reported effective masses of 0.35 $m_0$ and 0.31 $m_0$ for electrons and holes, respectively.[7] As a representative, the crystal structure, band structure and partial density of states (DOS) of BiSeCl are shown in Fig. 4(b). The crystal structure has chains of atoms running along the *c*-axis. The band structure shows an indirect band gap with the conduction band minimum (CBM) located at Γ and valence band maximum (VBM) along the Γ-Z direction. The conduction bands are highly dispersive both parallel (Γ-Z direction) and perpendicular (Γ-X and Γ-Y direction) to the chain direction ($k_z$), since they are mainly derived from the spatially extended Bi-*p* states. The valence bands are dispersive in direction parallel to the atomic chains (Γ-Z direction) but less so perpendicular (Γ-X and Γ-Y direction) to the chain. This is due to the weak Se-Se inter-chain coupling compared with that of intra-chain coupling. The shortest inter-chain Se-Se distance is 4.32 Å, which is larger than intra-chain Se-Se distance of 3.81 Å. We note that a prior density functional investigation[10] also indicated that BiSeBr and BiSeI may have potential as a solar absorber materials consistent with the present results. The band structures are given in Supplementary Fig. S2. Within these three compounds the halogen *p* states take an increasing role in the VBM as the halogen atomic number increases from Cl to I. For BiSeI, the nearest inter-chain I-I distance is slightly shorter than the intra-chain distances, indicating a stronger inter-chain I-I coupling. This is reflected in the band structure. Thus, the valence bands of BiSeI are dispersive in direction both parallel and perpendicular to the atomic chains (Supplementary Fig. S2).

$Bi_3Se_4Br$ is another candidate absorber identified in the present calculations. In contrast to the three compounds discussed above, it possesses simultaneously low electron and hole effective masses of 0.71 and 1.20 $m_0$, respectively, indicating efficient transport of both electrons and holes. The crystal structure of $Bi_3Se_4Br$ is depicted in the insert of Fig. 4(c). It also has atomic chains, in this case along the *b* axis. The calculated band structure shown in Fig. 4(c) indicates an indirect band gap (~1.69 eV) with the CBM at *k* point along the Γ-M direction and VBM along the N-Z direction. As expected, not only the valance bands but also the conduction bands are dispersive. The partial DOS [Fig. 4(c)] shows that the valence band is dominated by anion *p* states, as expected. These hybridize with Bi 6*s* orbital to form antibonding states, responsible for the dispersive valence bands. The conduction band is primarily derived from Bi 6*p* states.

We note that due to matrix elements and selection rules the magnitude of the electronic gap is not always indicative of strong absorption that is needed for solar applications, and calculations of absorption spectra are needed to assess their potential use in solar cells. The calculated absorption spectra for the four potential solar absorbers show strong absorptions near the band gap (Fig. 4(d); note that these are direct (vertical) transitions so absorption onset is slightly above the indirect gap), consistent with high solar-energy capture efficiency.

**Potential semiconductors for room-temperature radiation detection**

Seven (BiSCl, BiSBr, BiSI, BiOI, $Bi_3Se_4Br$, SbSI and SbSeI) of the 36 compounds show optical band gaps in range between 1.5-2.2 eV as well as electron or hole effective masses lower than $m_0$. These are then potentially useful as room-temperature radiation detection materials as far as this screen is concerned. Since heavy atoms (high *Z*) are necessary for efficient absorption of high energy radiation, five Bi based compounds are favorable, whose band gaps and carrier effective masses are shown in Fig. 5(a). Among the five compounds, $Bi_3Se_4Br$ has been discussed above.

BiSCl, BiSBr and BiSI adopt the same chain-like crystal structure as BiSeCl/Br/I with space group *Pnma* (No. 62). The band structure and partial DOS of BiSCl are shown in Fig. 5(b) as a representative, while those of BiSBr and BiSI are given in Supplementary Fig. S3. Generally, the band structures of BiSCl/Br/I are similar to those of BiSeCl/Br/I, except that the former have larger band gaps. This is attributable to the deeper S-*p* states as compared with Se-*p* states (recall the chalcogen *p* VBM character). Within this group of compounds, the valence bands are mainly composed of anion *p* states (S-*p* and halogen-*p*). The contributions from halogen-*p* states near the VBM increase as the halogen ion changes from Cl to I. In addition, there are also antibonding contributions from the Bi-*s* states. Despite this, the valence bands are still relatively flat leading to large hole effective masses, typically larger than $m_0$. The conduction bands are always dominated by the spatially extended Bi-*p* states, leading to low electron effective masses, less than $m_0$. The disparity between electron and hole effective mass may not be a prohibitive condition for radiation detection applications, as these devices are often designed to collect only one type of charge carrier.

BiOI is another candidate for radiation detection identified in this study. It crystallizes to a layered tetragonal structure of space group *P*4/*nmm* (No. 129). As depicted in the insert of Fig. 5(c), each Bi is coordinated by four O atoms, while each O has four 4 Bi neighbors, forming Bi-O layers. Each Bi-O layer is sandwiched by two iodine layers, stacking along the *c*-axis. The band structure of BiOI is shown in Fig. 5(c). It has an indirect band gap. The CBM is at Γ point, while the VBM is at a k point along the Z-R direction. The states near VBM are dominated by I-*p* states, while the states near CBM mainly derive from Bi-*p* states [Fig. 5(c)]. The inter- and intra-layer I-I distances are 4.89 and 4.03 Å, indicating a weak inter-layer interaction. This is responsible for the rather flat valence bands perpendicular to the atomic layers (Γ-Z and X-R directions). However, both valence and conduction bands along the Z-R are dispersive due to the strong interactions within the atomic layers. Remarkably, it possesses simultaneously low electron and hole effective masses, which raises the possibility of high mobility-lifetime ($\mu\tau$) product for both type of charge carriers – a desirable characteristic for efficient charge collection and energy resolution in semiconductor detectors.

It is known that high dielectric constants are favorable for long carrier lifetime, large diffusion length, and the higher $\mu\tau$ values via effective screening of charged defects. The calculated static dielectric constants ($\varepsilon_{st}$, the high-frequency limit resulted from electronic polarization) for the five potential radiation detection materials are shown in Fig. 5(d) (see also Fig. 3; numerical values in Table S3). As seen all the materials exhibit rather large $\varepsilon_{st}$ above 30, in spite of their sizable band gaps.

**Compounds as potential transparent conductors**

For transparent conducting materials, we focus on finding *p*-type (hole conducting) examples, since these are rather scarce compared to *n*-type.[37-41] In addition to large band gap (> 3 eV), low hole effective mass is a prerequisite since good conductivity is needed. While not reflecting a strictly intrinsic property, it should also be emphasized that successful development of a transparent conductor requires doping. This is a necessary criterion for a transparent conductor. The difficulty in doping most oxides p-type has been an important challenge in this field.[38] It has been overcome through specific chemistries, such has that of monovalent Cu and divalent Sn, and through particularly stable crystal structures such as that of $BaSnO_3$.[37-40] Five compounds (BiOCl, BiOBr, SbOF, $Sb_4O_5Cl_2$ and $Sb_4O_5Br_2$, Fig. 6) are found to possess optical band gaps larger than 3 eV as well as reasonable hole effective masses (below 1.5 $m_0$ expect for BiOBr with ~1.7 $m_0$). These may be potential transparent conducting materials, if they can be doped. All of them contain oxygen as the chalcogen, reflecting the fact that the high electronegativity of O is favorable for a large band gap.

BiOCl and BiOBr possess the same crystal structure of space group *P*4/*nmm* (No. 129) as that of BiOI. The band structures and partial DOS of BiOCl are shown in Fig. 6(b), and those of BiOBr are shown in Supplementary Fig. S4. Their band structures display similar features as that of BiOI, partially reflecting the fact that these compounds adopt the same crystal structure. However, the band gap increases as the halogen ion changes from I to Cl because of the lowering of the halogen *p* states, following the electronegativity. Moreover, the dispersion of the relatively flat conduction bands along Γ-Z (perpendicular to the atomic layers) also changes when going from I to Cl, leading to a shift in the CBM from Γ to Z point. The VBMs of the three compounds are always along the Z-R direction (parallel to the atomic layers). Due the strong intra-layer halogen-halogen coupling, the valence bands are highly dispersive along Z-R direction, leading to low hole effective masses.

SbOF has a chain-like structure with space group *Pnma* (No. 62). The atomic chains extend along the *b* axis [inset of Fig. 6(c)]. Although it has the same symmetry as the BiSeCl/Br/I and BiSCl/Br/I compounds discussed earlier, the bonding pattern in SbOF is distinct. The Sb-O bonds in SbOF are almost in the plane of the atomic chains, while the Bi-Se/S bonds are not. The band structure and partial DOS of SbOF are shown in Fig. 6(c). It has an indirect band gap of 3.95 eV with VBM at Γ and CBM at Y. The VBM is dominated by O-2*p* states. Since the nearest inter-chain O-O distance (4.77 Å) is much larger than that of intra-chain distance (2.47 Å), the inter-chain O-O coupling is rather weak. This is responsible for the flat valance bands in direction perpendicular to the chain (Γ-X). However, the valence bands are highly dispersive along the chain direction (Γ-Y) due to the strong intra-chain interactions.

$Sb_4O_5Cl_2$ and $Sb_4O_5Br_2$ adopt same layered structure with space group $P2_1/c$ (No. 14) and atomic layers stacked along *a* axis. Among these, $Sb_4O_5Br_2$ was previously suggested as *p*-type transparent conducting material based on density functional calculations.[39] The band structures of the two compounds shown in Fig. 6(d) ($Sb_4O_5Br_2$) and Supplementary Fig. S5 ($Sb_4O_5Cl_2$) are very similar. Both have indirect band gaps, VBM at Γ and CBM at Z. Interestingly, although they have a layered structure, the valence bands are dispersive in direction perpendicular to the atomic layers (Γ-Z direction). As seen in the DOS [Fig. 6(d)], the states near the VBM are dominated by halogen *p* states. The intra-layer halogen-halogen distances (5.10 and 5.13 Å for $Sb_4O_5Cl_2$ and $Sb_4O_5Br_2$, respectively) are much larger than those of interlayer ones (4.47 and 4.39 Å for $Sb_4O_5Cl_2$ and $Sb_4O_5Br_2$, respectively). This leads to strong interlayer halogen-halogen interactions and explains the dispersive valence bands in the Γ-Z direction.

## DISCUSSION

Motivated by recent results on halide semiconductors, and theoretical arguments about the origin of their performance, we selected a group of 36 chalcohalides and oxyhalides containing n$s^2$ $Bi^{3+}$ or $Sb^{3+}$ cation. Density functional results for band gap, effective mass of electrons and holes, and optical absorption are reported. BiSeCl/Br/I and $Bi_3Se_4Br$ are found to possess optical band gaps of 1.47-1.69 eV and low electron or hole effective masses. They are promising solar absorber materials. $Bi_3Se_4Br$, BiSCl/Br/I and BiOI have band gaps in the range 1.69-2.08 eV, and low effective mass of electrons or holes, suitable for use as room-temperature radiation detection materials. BiOCl/Br, SbOF, and $Sb_4O_5Cl_2/Br_2$ have larger band gaps (>3.0 eV) and low hole effective mass, which makes them potential candidates as *p*-type transparent conducting materials. This suggests further study of the identified materials to assess other properties relevant to applications, such as dopability in the case of transparent conductors, and the possibility for obtaining high resistivity in the case of semiconductors for radiation detection.

There are several common themes in the electronic structure of the identified compounds. The band gap is controlled by positions of the anion *p*-states which form the valence bands. The Bi or Sb *s*-states hybridize in antibonding combinations with halogen-*p* and chalcogen-*p* states near respective VBMs favoring high dispersion and lower hole effective mass. The CBM of the compounds is formed primarily from Bi or Sb *p*-states. There is also a tendency towards cross-gap hybridization between the cation *p*- and anion *p*-states (band structure and DOS Figs. 3, 4, and 5). Cross-gap hybridization indicates covalency in the otherwise ionic materials, and possibility of large Born effective charge and static dielectric constant. The identified compounds do as a result have high dielectric constants. These large static dielectric constants are favorable for effective screening of charged defects and high mobility. Despite the complex chain or layered structures of these compounds, the valence and conduction band dispersion is a consequence of hybridization among the anion or cation *p*-states, favoring low carrier effective mass. The current work proposes a promising series of Bi and Sb based oxyhalides and chalcohalides, which may be appropriate for various optoelectronic applications. Further theoretical and experimental investigations will be necessary to ascertain their dopability, defect and dielectric behavior, the possibility of obtaining high resistivity for radiation detection, as well as device-level properties.

**METHODS**

All calculations are performed within the framework of density functional theory (DFT) using plane-wave pseudopotential method as implemented in the VASP code.[42] The electron-ion interaction is described by means of projector-augmented wave pseudopotentials.[42,43] Valence electron configurations of n$s^2$n$p^3$ for Sb/Bi, n$s^2$n$p^4$ O/S/Se/Te and n$s^2$n$p^5$ for F/Cl/Br/I are employed along with the Perdew-Burke-Ernzerhof generalized gradient approximation (PBE-GGA).[44] A plane-wave basis set with an energy cutoff of 500 eV and the Monkhorst-Pack *k*-point mesh with grid spacing of ~$2\pi\times$ 0.03 Å$^{-1}$ are used for total energy minimization and ground-state electronic structure calculations. Once the basic properties were obtained in this way, we then calculated electronic structures using the hybrid Heyd-Scuseria-Ernzerhof (HSE)[45] functional, with the standard 25% non-local Fock exchange. Spin–orbit coupling (SOC) is included. This is important for these heavy *p*-electron element containing compounds. The carrier effective mass ($m^*$) tensors for transport are directly calculated using the BoltzTraP code.[46] For this purpose, the PBE+SOC eigenvalues on a dense k-point grid ($2\pi\times$0.015 Å$^{-1}$ or less), with a carrier concentration of $10^{18}$ cm$^{-3}$ and temperature of 300 K. This takes into account effects of non-parabolicity, anisotropy of bands, multiple bands, and other band structure effects, to obtain a single effective mass tensor for carrier transport. The quoted average values are from the direction average of the inverse transport effective mass.

## ACKNOWLEDGEMENTS

The authors acknowledge funding support from the National Natural Science Foundation of China (under Grant Nos. 61722403 and 11674121), National Key Research and Development Program of China (under Grants No. 2016YFB0201204), Program for JLU Science and Technology Innovative Research Team, and the Recruitment Program of Global Youth Experts in China. Work at the University of Missouri was supported by the Department of Energy through the S3TEC Energy Frontier Research Center, award DE-SC0001299. KB acknowledges support from US Department of Homeland Security under Grant Award Number, 2014-DN-077-ARI075. Part of the calculations was performed at the High Performance Computing Center of Jilin University.

## COMPETING INTERESTS

We declare that we have no competing interests in regard to this work.

## AUTHOR CONTRIBUTIONS

XR, XW, YL, DY and XGZ performed calculations. KB contributed to the analysis and discussion for radiation detection materials. DJS and LZ analysed results and wrote the draft. All authors contributed to writing the final manuscript. XR and XW contributed equally.

**Table I.** Calculated and experimental band gaps (Refs. 18, 20-34) of some representative compounds considered in the current work.

| Material | Band gap(eV) | | |
|---|---|---|---|
| | HSE+SOC | | Exp. |
| | Direct | Indirect | |
| BiOF | 3.94 | - | 4.01[20] |
| BiOCl | 3.80 | 3.29 | 3.32,[20] 3.20,[21] 3 .44[22] |
| $Sb_4O_5Cl_2$ | 3.95 | 3.50 | 3.30[23] |
| BiOBr | 3.09 | 2.86 | 2.85,[24] 2.76[22] |
| BiOI | 2.08 | 1.90 | 1.92,[25] 1.85[22] |
| BiSCl | 1.98 | - | 1.93[26] |
| BiSBr | 1.89 | - | 1.95[27] |
| BiSI | 1.87 | 1.66 | 1.57,[18] 1.592[28] |
| BiSeBr | 1.48 | 1.45 | 1.54[29] |
| BiSeI | 1.48 | 1.32 | 1.29,[30] 1.32,[29] 1.285[28] |
| BiTeCl | 1.33 | 1.20 | 0.77,[31] ~0.8[32] |
| BiTeI | 1.09 | 0.90 | 0.38[33] |
| SbTeI | 1.29 | 1.20 | 0.73,[34] 1.28[26] |

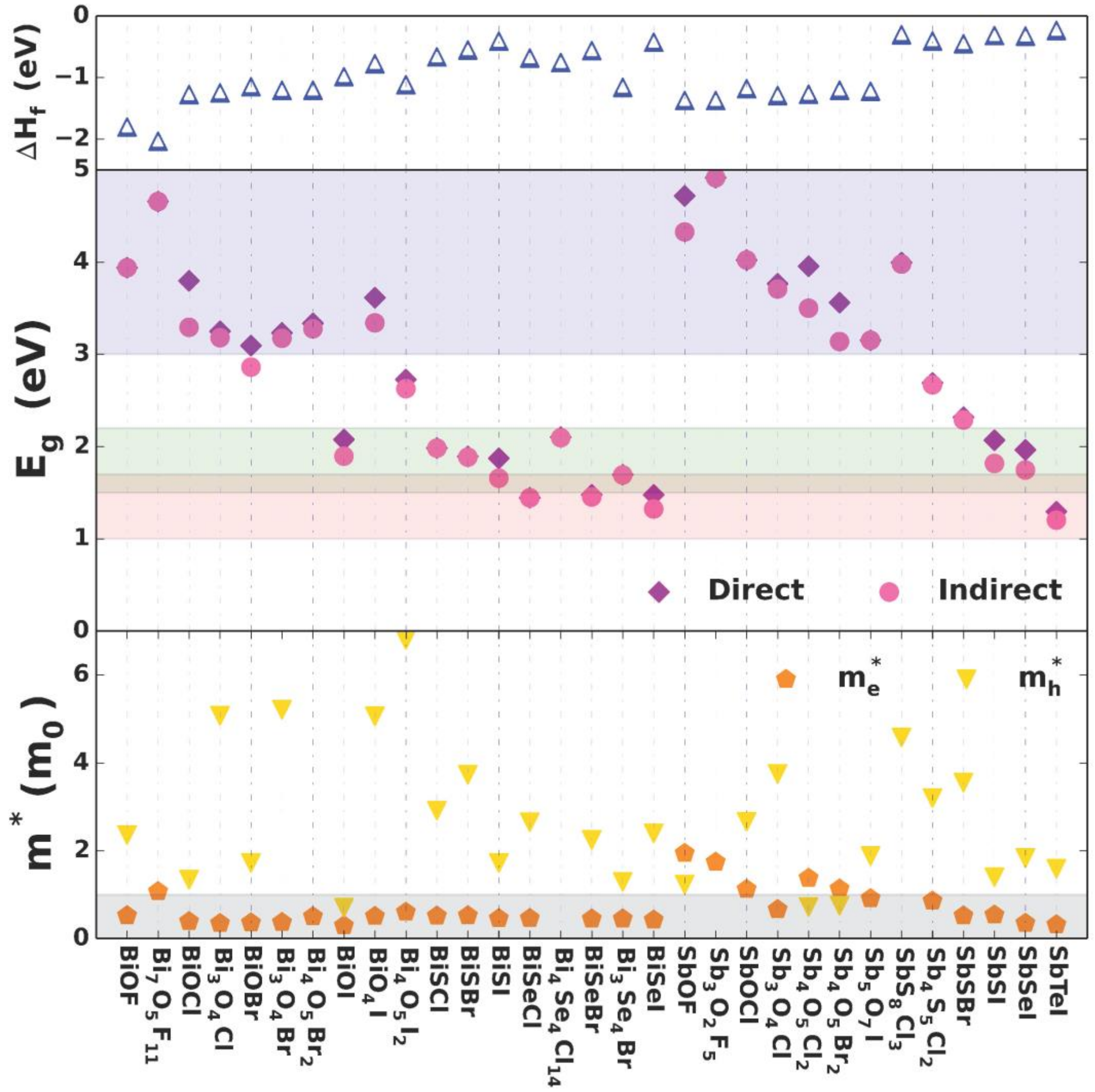

**Figure 1.** Formation energies per atom (upper panel), band gaps (middle panel) and effective masses of electron and hole (lower panel) for the 31 Bi and Sb based compounds. The formation energy is defined as energy difference between the compound and constituent elemental solids. Red, green and blue shaded areas in the middle panel indicate band gap ranges of 1.0-1.7, 1.5-2.2 and >3 eV, respectively. Gray shaded area in the lower panel indicates effective masses lower than the rest mass of electron ($m_0$).

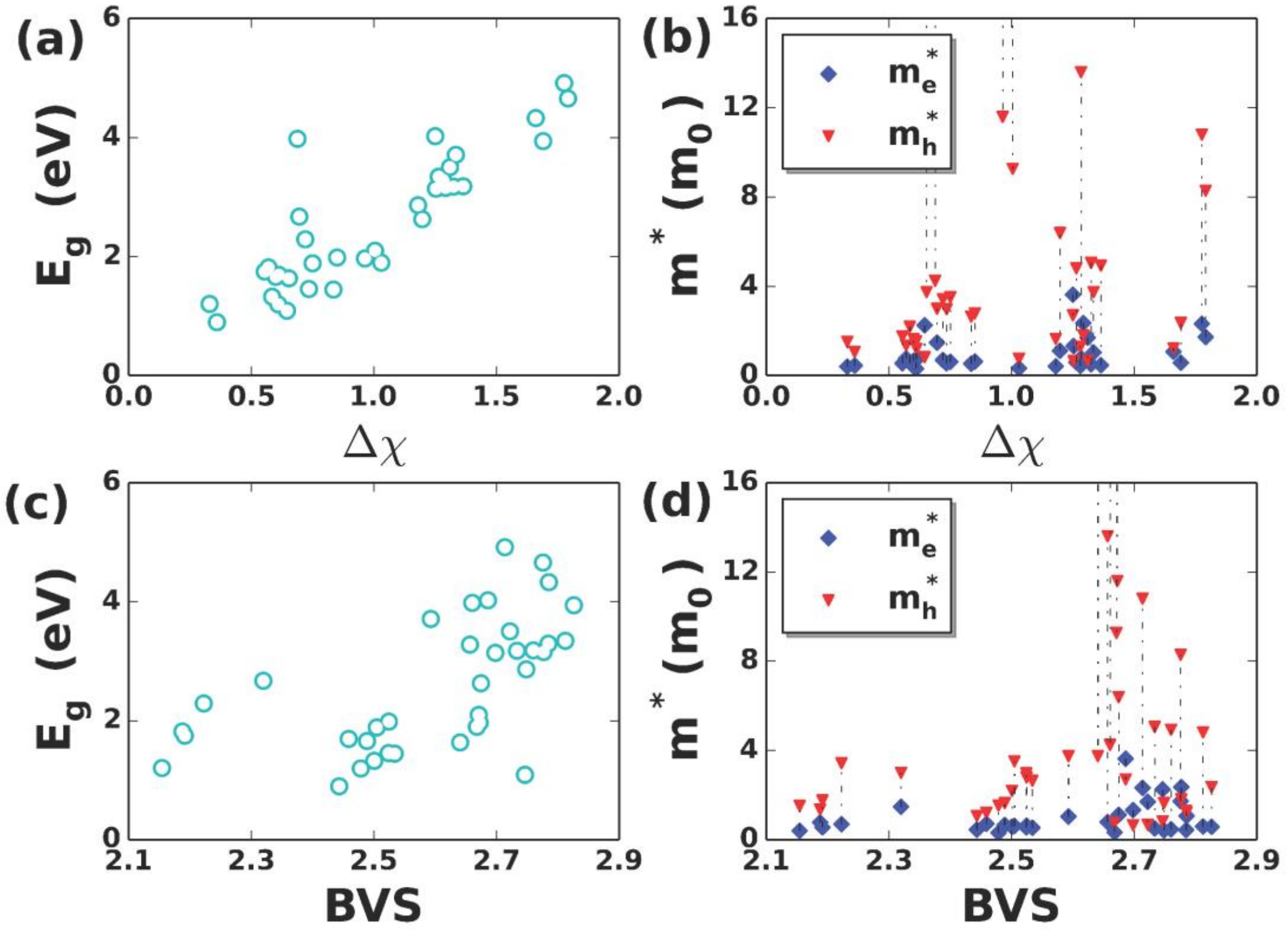

**Figure 2.** (a) Band gaps, (b) hole and electron effective masses as a function of average electronegativity discrepancy between anions and cations for the 31 Bi and Sb based compounds; (c) Band gaps, (d) effective masses of electron and hole as a function of bond valence sum of the cations.

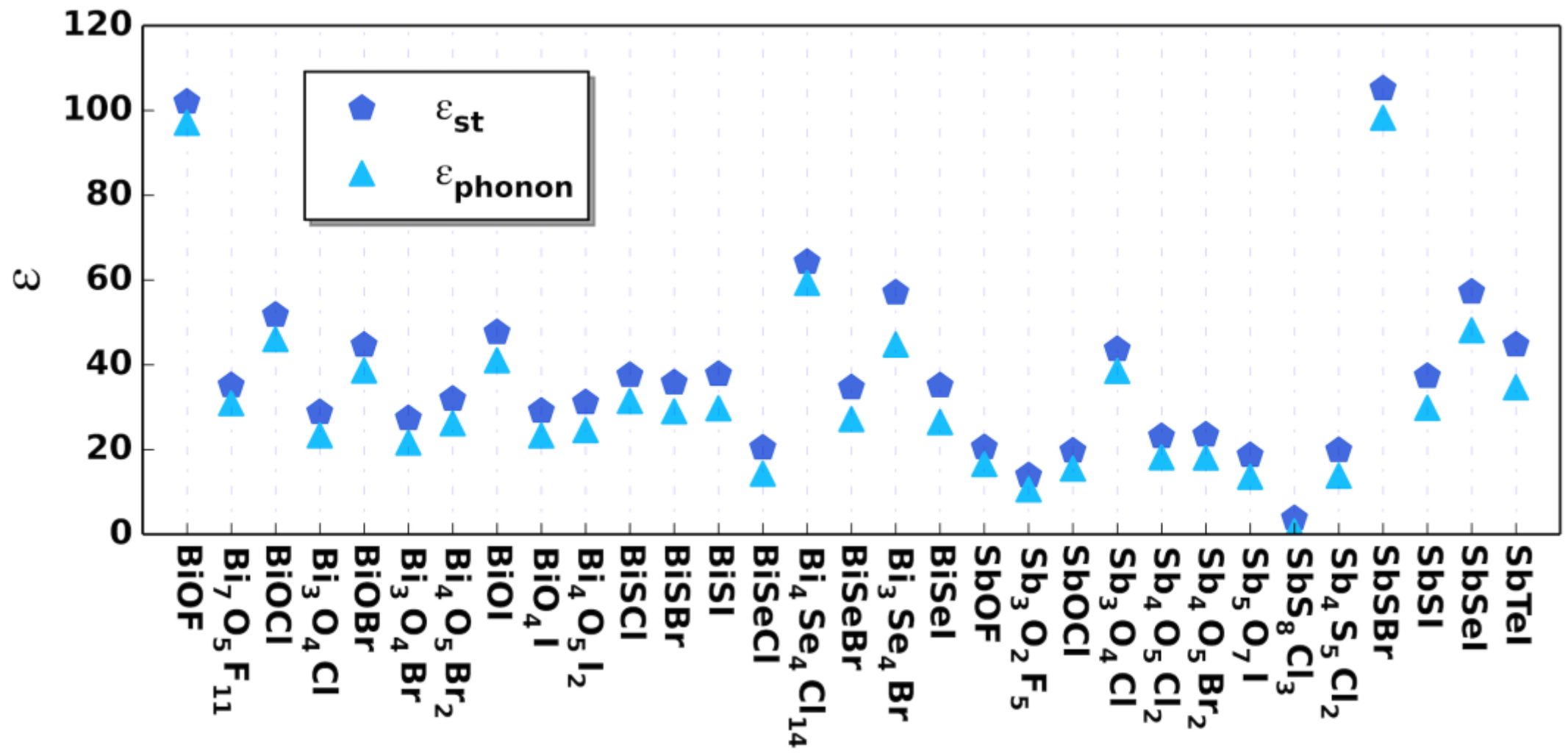

**Figure 3.** Calculated dielectric constants (ε) of the considered bismuth/antimony oxyhalides and chalcohalides. We show total static dielectric constant ($\varepsilon_{st}$) and the component from phonon contribution ($\varepsilon_{phonon}$). Note that these compounds have generally high dielectric constants, mostly due to large phonon contributions.

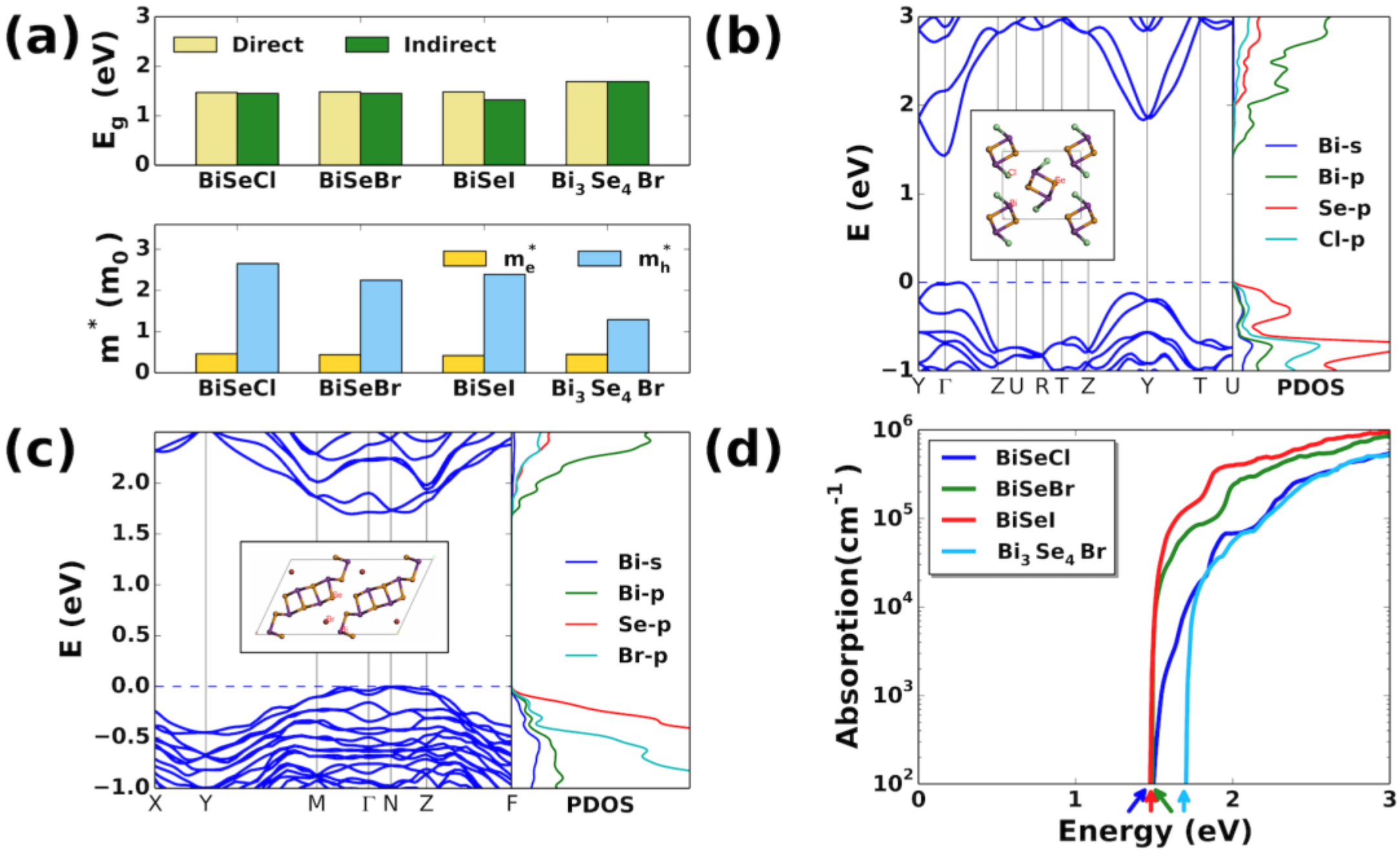

**Figure 4.** (a) Histogram of HSE+SOC band gaps, effective masses of electron and hole for BiSeCl, BiSeBr, BiSeI and $Bi_3Se_4Br$. (b) Crystal structure, band structure and partial density of states (DOS) of BiSeCl. (c) Crystal structure, band structure and partial DOS of $Bi_3Se_4Br$. (d) Absorption spectra of BiSeCl, BiSeBr, BiSeI and $Bi_3Se_4Br$.

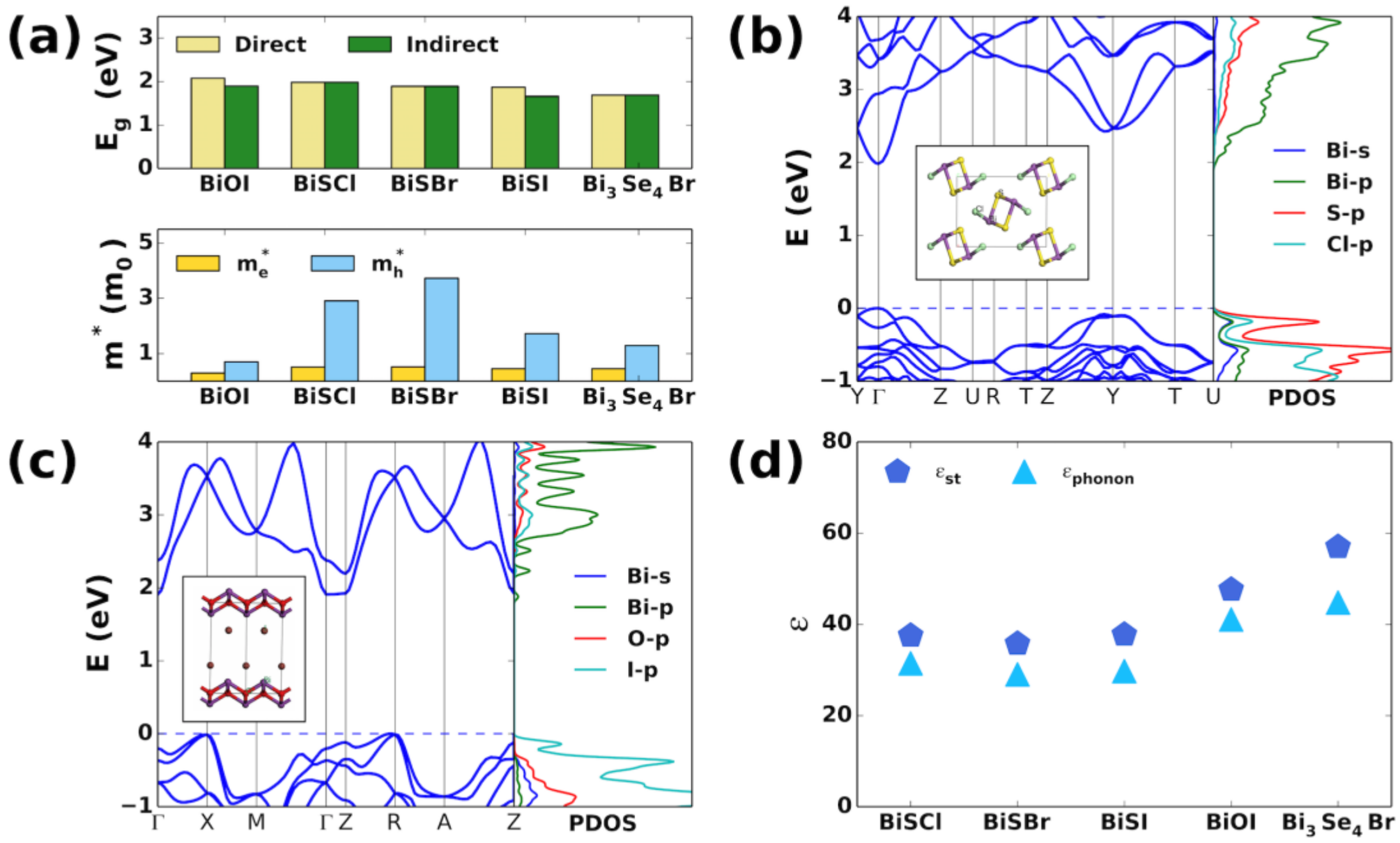

**Figure 5.** (a) Histogram of HSE+SOC band gaps, effective masses of electron and hole for BiOI, BiSCl, BiSBr, BiSI and $Bi_3Se_4Br$. (b) Crystal structure, band structure and partial DOS of BiSCl. (c) Crystal structure, band structure and partial DOS of BiOI. (d) Phonon contribution to, and total static dielectric constant ($\varepsilon_{st} = \varepsilon_{\infty}+\varepsilon_{phonon}$) for BiSCl, BiSBr, BiSI, BiOI, and $Bi_3Se_4Br$.

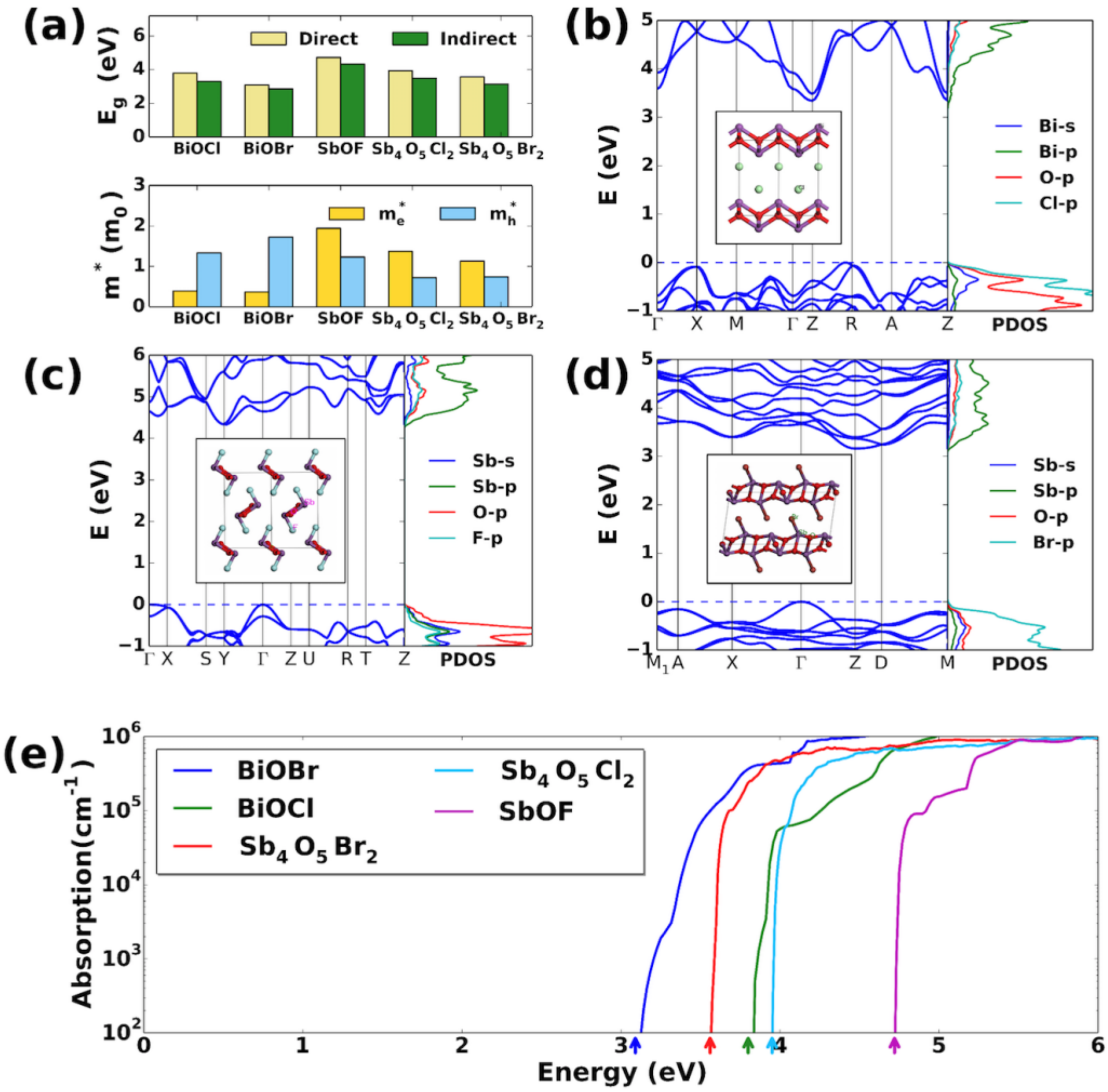

**Figure 6.** (a) Histogram of HSE+SOC band gaps, effective masses of electron and hole for BiOCl, BiOBr, SbOF, $Sb_4O_5Cl_2$ and $Sb_4O_5Br_2$. (b) Crystal structure, band structures and partial DOS of BiOCl. (c) Crystal structure, band structure and partial DOS of SbOF. (d) Crystal structure, band structure and partial DOS of $Sb_4O_5Br_2$. (e) Calculated absorption spectra of BiOBr, BiOCl, $Sb_4O_5Br_2$, $Sb_4O_5Cl_2$ and SbOF.

# Supplementary Information

## *Bismuth and antimony-based oxyhalides and chalcohalides as potential optoelectronic materials*

Zhao Ran[1,†], Xinjiang Wang[1,†], Yuwei Li[2], Dongwen Yang[1], Xin-Gang Zhao[1], Koushik Biswas[3], David J. Singh[2,*], and Lijun Zhang[1,*]

[1]*Key Laboratory of Automobile Materials of MOE, State Key Laboratory of Superhard Materials, and College of Materials Science, Jilin University, Changchun 130012, China*

[2]*Department of Physics and Astronomy, University of Missouri, Columbia, MO 65211-7010 USA*

[3]*Department of Chemistry and Physics, Arkansas State University, State University, AR 72467, USA*

[†]These authors contributed equally.

[*]Address correspondence to: singhdj@missouri.edu or lijun_zhang@jlu.edu.cn

**Table S1.** Space group, electronic band gaps, effective masses of carriers and formation energy $\Delta H_f$ of the 31 bismuth/antimony oxyhalides and chalcohalides considered in the current work. For the effective masses, the "*" means a very large number.

| | *Compounds* | *Space Group* | *Band Gap* (eV) | | | | *Effective masses* ($m_0$) | | | | | | | | $\Delta H_f$ (eV) |
|---|---|---|---|---|---|---|---|---|---|---|---|---|---|---|---|
| | | | PBE | PBE+SOC | HSE+SOC | | electron | | | | hole | | | | |
| | | | | | Direct | Indirect | $m^*$ | $m_{xx}$ | $m_{yy}$ | $m_{zz}$ | $m^*$ | $m_{xx}$ | $m_{yy}$ | $m_{zz}$ | |
| 1 | BiOF | *P4/nmm* | 3.13 | 2.77 | 3.94 | - | 0.52 | 0.42 | 0.42 | 1.04 | 2.35 | 5.09 | 5.09 | 1.13 | -1.87 |
| 2 | $Bi_7O_5F_{11}$ | *C2* | 3.89 | 3.44 | 4.66 | - | 1.07 | 0.95 | 0.72 | 2.70 | 9.24 | * | * | 3.75 | -2.10 |
| 3 | BiOCl | *P4/nmm* | 2.56 | 2.34 | 3.80 | 3.29 | 0.39 | 0.29 | 0.29 | 1.15 | 1.33 | 0.95 | 0.95 | 6.31 | -1.31 |
| 4 | Bi3O4Cl | *C2/c* | 2.42 | 2.27 | 3.25 | 3.18 | 0.34 | 0.29 | 0.34 | 0.42 | 5.07 | 5.43 | 2.49 | * | -1.30 |
| 5 | BiOBr | *P4/nmm* | 2.25 | 2.00 | 3.09 | 2.86 | 0.36 | 0.24 | 0.24 | * | 1.72 | 1.26 | 1.26 | 6.15 | -1.19 |
| 6 | $Bi_3O_4Br$ | *Pnna* | 2.29 | 2.14 | 3.23 | 3.17 | 0.36 | 0.30 | 0.33 | 0.52 | 5.21 | 5.65 | 2.53 | * | -1.26 |
| 7 | $Bi_4O_5Br_2$ | *P2₁* | 2.51 | 2.31 | 3.34 | 3.28 | 0.50 | 1.64 | 0.33 | 0.70 | * | * | * | * | -1.25 |
| 8 | BiOI | *P4/nmm* | 1.46 | 1.21 | 2.08 | 1.90 | 0.29 | 0.19 | 0.19 | * | 0.70 | 0.51 | 0.51 | * | -1.02 |
| 9 | $BiO_4I$ | *Pca2₁* | 2.38 | 2.22 | 3.61 | 3.34 | 0.50 | 0.45 | 1.43 | 0.33 | 5.00 | 3.12 | * | 4.46 | -0.85 |
| 10 | $Bi_4O_5I_2$ | *P2₁* | 2.24 | 1.82 | 2.73 | 2.63 | 0.60 | 1.15 | 0.35 | 0.79 | 6.81 | * | 2.59 | * | -1.16 |
| 11 | BiSCl | *Pnma* | 1.88 | 1.27 | 1.98 | - | 0.51 | 1.39 | 0.59 | 0.29 | 2.91 | 7.07 | 4.46 | 1.51 | -0.89 |
| 12 | BiSBr | *Pnma* | 1.84 | 1.21 | 1.89 | - | 0.52 | 1.30 | 0.85 | 0.26 | 3.73 | 7.76 | 4.97 | 2.11 | -0.77 |
| 13 | BiSI | *Pnma* | 1.87 | 1.07 | 1.87 | 1.66 | 0.45 | 1.87 | 0.49 | 0.24 | 1.72 | 3.17 | 1.53 | 1.29 | -0.63 |
| 14 | BiSeCl | *Pnma* | 1.53 | 0.87 | 1.47 | 1.45 | 0.46 | 0.54 | 0.99 | 0.27 | 2.65 | 2.47 | 2.44 | 3.12 | -0.68 |
| 15 | $Bi_4Se_4Cl_{14}$ | *P4/n* | 1.39 | 1.37 | 2.10 | - | * | * | * | * | * | 7.43 | 7.43 | * | -0.76 |
| 16 | BiSeBr | *Pnma* | 1.59 | 0.87 | 1.48 | 1.45 | 0.44 | 1.15 | 0.60 | 0.24 | 2.25 | 7.82 | 3.38 | 1.98 | -0.56 |
| 17 | $Bi_3Se_4Br$ | *C2/m* | 1.06 | 0.67 | 1.69 | - | 0.45 | 1.58 | 0.20 | 1.02 | 1.29 | 2.03 | 0.75 | 2.01 | -0.43 |
| 18 | BiSeI | *Pnma* | 1.57 | 0.82 | 1.48 | 1.32 | 0.42 | 1.06 | 0.56 | 0.22 | 2.39 | 4.38 | 1.89 | 2.00 | -0.43 |
| 19 | SbOF | *Pnma* | 3.28 | 3.25 | 4.72 | 4.33 | 1.94 | 0.74 | 0.74 | 2.01 | 1.23 | 7.57 | 0.78 | 0.97 | -1.86 |
| 20 | $Sb_3O_2F_5$ | *P2/c* | 3.80 | 3.76 | 4.91 | - | 1.74 | 9.63 | 1.20 | 1.26 | * | * | 5.95 | * | -2.07 |
| 21 | SbOCl | *P2₁/C* | 3.09 | 3.00 | 4.02 | - | 1.10 | 2.70 | 2.67 | 0.51 | 2.66 | 1.93 | * | 4.15 | -1.22 |
| 22 | $Sb_3O_4Cl$ | *P2/c* | 2.68 | 2.65 | 3.76 | 3.71 | 0.66 | 1.03 | 0.62 | 0.51 | 3.74 | 2.53 | 4.87 | 4.98 | -1.34 |
| 23 | $Sb_4O_5Cl_2$ | *P2₁/C* | 2.64 | 2.56 | 3.95 | 3.50 | 1.37 | 0.88 | 1.60 | 2.33 | 0.72 | 0.67 | 0.72 | 0.78 | -1.32 |
| 24 | $Sb_4O_5Br_2$ | *P2₁/C* | 2.38 | 2.27 | 3.56 | 3.14 | 1.13 | 0.72 | 1.01 | 3.53 | 0.74 | 0.74 | 0.69 | 0.82 | -1.25 |
| 25 | $Sb_5O_7I$ | *P-62c* | 2.45 | 2.26 | 3.15 | - | 0.91 | 0.76 | 0.80 | 1.27 | 1.88 | * | * | 0.70 | -1.28 |
| 26 | $SbS_8Cl_3$ | *P-1* | 2.88 | 2.87 | 3.99 | 3.98 | 8.39 | 6.00 | * | 9.65 | 4.58 | * | 7.64 | 2.11 | -1.05 |
| 27 | $Sb_4S_5Cl_2$ | *Pnma* | 1.91 | 1.86 | 2.69 | 2.67 | 0.85 | 1.58 | 0.47 | 1.28 | 3.19 | 2.84 | 5.64 | 2.43 | -0.71 |
| 28 | SbSBr | *Pnma* | 1.72 | 1.57 | 2.32 | 2.29 | 0.52 | 1.55 | 0.71 | 0.27 | 3.55 | 5.49 | 3.93 | 2.45 | -0.67 |
| 29 | SbSI | *Pnma* | 1.53 | 1.26 | 2.07 | 1.82 | 0.53 | 3.18 | 0.64 | 0.26 | 1.39 | 1.27 | 2.41 | 1.05 | -0.54 |
| 30 | SbSeI | *Pnma* | 1.38 | 1.16 | 1.96 | 1.75 | 0.35 | 0.93 | 0.36 | 0.21 | 1.83 | 1.89 | 1.78 | 1.83 | -0.33 |
| 31 | SbTeI | *C2/m* | 0.87 | 0.73 | 1.29 | 1.20 | 0.31 | 1.58 | 0.19 | 0.27 | 1.59 | 2.61 | 1.66 | 1.11 | -0.23 |

**Table S2:** Stability calculation for the formation energy of BiSeCl/Br/I and $Bi_3Se_4Br$ with structural optimization including van der Waals interactions based on the optB86 functional.

| Compound | Energy(eV) | Path | $\Delta H_f$(eV/atom) |
|---|---|---|---|
| **BiSeCl** | -11.22 | | |
| Bi | -3.87 | i. BiSeCl —> Bi + Se +Cl | -0.70 |
| Se | -3.45 | | |
| Cl | -1.80 | ii. 3BiSeCl —> $Bi_2Se_3$ + $BiCl_3$ | -0.15 |
| $SeCl_4$ | -12.45 | | |
| $BiCl_3$ | -13.05 | iii. 4BiSeCl —> $SeCl_4$ + $Bi_2Se_3$ + 2Bi | -1.51 |
| $Bi_2Se_3$ | -20.15 | | |
| **BiSeBr** | -10.63 | | |
| Bi | -3.87 | i. BiSeBr —> Bi + Se + Br | -0.57 |
| Se | -3.45 | | |
| Br | -1.59 | ii. 3BiSeBr —> $Bi_2Se_3$ + $BiBr_3$ | -0.08 |
| $Bi_2Se_3$ | -20.15 | | |
| SeBr | -5.24 | iii. BiSeBr —> SeBr + Bi | -0.51 |
| $BiBr_3$ | -11.53 | | |
| **BiSeI** | -10.11 | | |
| Bi | -3.87 | i. BiSeI —> Bi +Se +I | -0.44 |
| Se | -3.45 | | |
| I | -1.49 | ii. 3BiSeI —> $Bi_2Se_3$ + $BiI_3$ | +0.04 |
| $Bi_2Se_3$ | -20.15 | | |
| $BiI_3$ | -10.29 | | |
| **$Bi_3Se_4Br$** | -30.68 | | |
| Bi | -3.87 | i. $Bi_3Se_4Br$ —> 3Bi + 4Se +Br | -0.46 |
| Se | -3.45 | | |
| Br | -1.59 | ii. $3Bi_3Se_4Br$ —> $4Bi_2Se_3$ + $BiBr_3$ | +0.01 |
| $Bi_2Se_3$ | -20.15 | | |
| $BiBr_3$ | -11.53 | iii. $Bi_3Se_4Br$ —> SeBr + $Bi_2Se_3$ + Bi | -0.18 |
| SeBr | -5.24 | | |

**Table S3.** Calculated total static dielectric constant ($\varepsilon_{st}$), the high-frequency limit of dielectric constant ($\varepsilon_{\infty}$) and the component from phonon contribution ($\varepsilon_{phonon}$) for the 31 bismuth/antimony oxyhalides and chalcohalides.

| | *Compounds* | $\varepsilon_{st}$ | | | | $\varepsilon_{\infty}$ | | | | $\varepsilon_{phonon}$ | | | |
|---|---|---|---|---|---|---|---|---|---|---|---|---|---|
| | | Averaged | xx | yy | zz | Averaged | xx | yy | zz | Averaged | xx | yy | zz |
| 1 | BiOF | 102.01 | 141.90 | 141.90 | 22.24 | 4.91 | 5.20 | 5.20 | 4.35 | 97.10 | 136.70 | 136.70 | 17.89 |
| 2 | $Bi_7O_5F_{11}$ | 35.13 | 34.87 | 17.32 | 53.19 | 4.17 | 4.20 | 3.84 | 4.52 | 30.94 | 30.67 | 13.48 | 48.67 |
| 3 | BiOCl | 51.74 | 73.10 | 73.10 | 9.01 | 5.69 | 6.30 | 6.30 | 4.47 | 46.04 | 66.80 | 66.80 | 4.54 |
| 4 | Bi3O4Cl | 28.76 | 36.05 | 24.72 | 25.52 | 5.43 | 5.63 | 5.38 | 5.28 | 23.34 | 30.42 | 19.34 | 20.24 |
| 5 | BiOBr | 44.62 | 63.58 | 63.58 | 6.70 | 5.98 | 6.78 | 6.78 | 4.40 | 38.64 | 56.80 | 56.80 | 2.30 |
| 6 | $Bi_3O_4Br$ | 27.32 | 35.79 | 21.73 | 24.44 | 5.69 | 5.86 | 5.66 | 5.55 | 21.63 | 29.93 | 16.07 | 18.89 |
| 7 | $Bi_4O_5Br_2$ | 31.92 | 31.31 | 37.91 | 26.54 | 5.72 | 5.74 | 5.90 | 5.57 | 26.18 | 25.57 | 32.01 | 20.97 |
| 8 | BiOI | 47.62 | 68.61 | 68.61 | 5.63 | 6.57 | 7.67 | 7.67 | 4.36 | 41.05 | 60.94 | 60.94 | 1.27 |
| 9 | $BiO_4I$ | 29.07 | 28.60 | 18.18 | 40.43 | 5.76 | 5.97 | 5.35 | 5.95 | 23.31 | 22.63 | 12.83 | 34.48 |
| 10 | $Bi_4O_5I_2$ | 31.09 | 33.08 | 31.75 | 28.44 | 6.52 | 6.53 | 6.68 | 6.37 | 24.56 | 26.55 | 25.07 | 22.07 |
| 11 | BiSCl | 37.53 | 9.10 | 22.83 | 80.66 | 6.08 | 4.53 | 5.96 | 7.77 | 31.44 | 4.57 | 16.87 | 72.89 |
| 12 | BiSBr | 35.75 | 9.22 | 22.08 | 75.94 | 6.75 | 5.12 | 6.54 | 8.59 | 29.00 | 4.10 | 15.54 | 67.35 |
| 13 | BiSI | 37.81 | 9.48 | 22.13 | 81.83 | 8.03 | 6.17 | 7.60 | 10.31 | 29.72 | 3.31 | 14.53 | 71.52 |
| 14 | BiSeCl | 20.38 | 19.48 | 9.34 | 32.31 | 6.08 | 5.72 | 4.76 | 7.75 | 14.30 | 13.76 | 4.58 | 24.56 |
| 15 | $Bi_4Se_4Cl_{14}$ | 64.20 | 92.24 | 92.24 | 8.13 | 4.91 | 5.51 | 5.51 | 3.69 | 59.30 | 86.73 | 86.73 | 4.44 |
| 16 | BiSeBr | 34.59 | 9.70 | 24.00 | 70.08 | 7.42 | 5.44 | 7.39 | 9.43 | 27.17 | 4.26 | 16.61 | 60.65 |
| 17 | $Bi_3Se_4Br$ | 56.98 | 38.54 | 92.03 | 40.38 | 12.21 | 11.31 | 14.59 | 10.74 | 44.77 | 27.23 | 77.44 | 29.64 |
| 18 | BiSeI | 35.05 | 10.14 | 71.71 | 23.30 | 8.6 | 6.47 | 10.95 | 8.36 | 26.45 | 3.67 | 60.76 | 14.94 |
| 19 | SbOF | 20.49 | 7.44 | 46.88 | 7.14 | 3.91 | 3.41 | 4.88 | 3.43 | 16.57 | 4.03 | 42.00 | 3.71 |
| 20 | $Sb_3O_2F_5$ | 13.84 | 21.26 | 8.73 | 11.54 | 3.25 | 3.36 | 3.12 | 3.29 | 10.59 | 17.90 | 5.61 | 8.25 |
| 21 | SbOCl | 19.68 | 16.65 | 16.82 | 25.56 | 4.18 | 4.21 | 4.02 | 4.32 | 15.49 | 12.44 | 12.80 | 21.24 |
| 22 | $Sb_3O_4Cl$ | 43.62 | 58.28 | 13.22 | 59.35 | 5.09 | 5.21 | 4.81 | 5.24 | 38.53 | 53.07 | 8.41 | 54.11 |
| 23 | $Sb_4O_5Cl_2$ | 23.09 | 22.19 | 18.89 | 28.18 | 4.92 | 4.78 | 5.05 | 4.94 | 18.16 | 17.41 | 13.84 | 23.24 |
| 24 | $Sb_4O_5Br_2$ | 23.45 | 25.57 | 17.86 | 26.91 | 5.37 | 5.40 | 5.51 | 5.20 | 18.08 | 20.17 | 12.35 | 21.71 |
| 25 | $Sb_5O_7I$ | 18.62 | 20.00 | 20.02 | 15.85 | 5.05 | 5.11 | 5.11 | 4.93 | 13.56 | 14.89 | 14.91 | 10.92 |
| 26 | $SbS_8Cl_3$ | 3.73 | 3.69 | 3.59 | 3.90 | 3.01 | 3.14 | 2.86 | 3.05 | 0.71 | 0.55 | 0.73 | 0.85 |
| 27 | $Sb_4S_5Cl_2$ | 19.78 | 16.53 | 36.96 | 5.86 | 5.97 | 5.88 | 7.57 | 4.45 | 13.82 | 10.65 | 29.39 | 1.41 |
| 28 | SbSBr | 105.15 | 7.33 | 15.53 | 292.6 | 6.86 | 4.96 | 6.28 | 9.35 | 98.27 | 2.37 | 9.25 | 283.25 |
| 29 | SbSI | 37.28 | 8.37 | 14.34 | 89.12 | 7.46 | 5.56 | 6.55 | 10.27 | 29.82 | 2.81 | 7.79 | 78.85 |
| 30 | SbSeI | 57.18 | 8.59 | 18.22 | 144.73 | 9.03 | 6.35 | 8.43 | 12.34 | 48.14 | 2.24 | 9.79 | 132.39 |
| 31 | SbTeI | 44.69 | 7.93 | 99.97 | 26.18 | 10.04 | 6.19 | 13.61 | 10.31 | 34.66 | 1.74 | 86.36 | 15.87 |

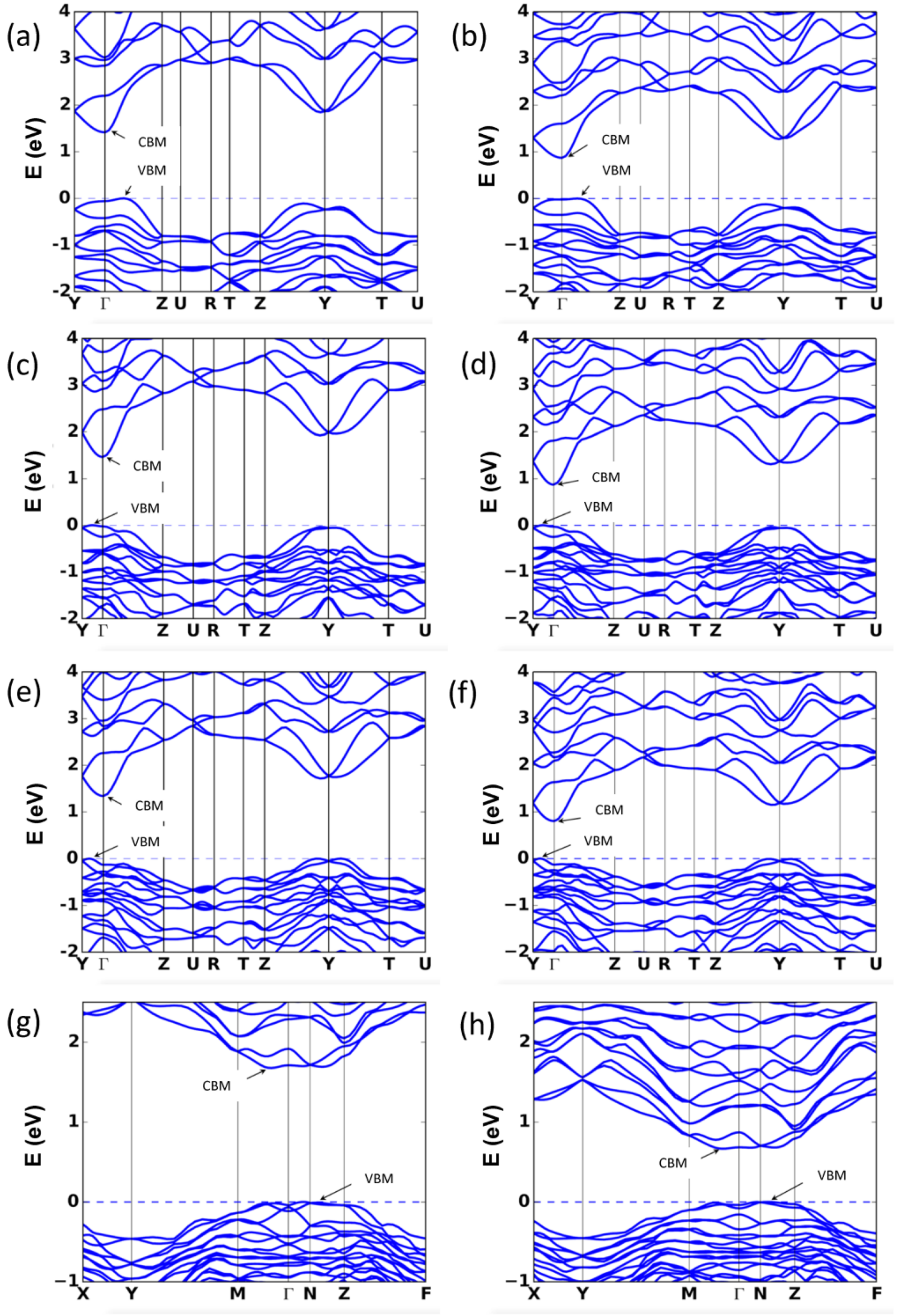

**Figure S1.** Band structure comparison of HSE+SOC (left) and PBE+SOC (right) calculations for BiSeCl (a, b), BiSeBr (c,d), BiSeI (e,f) and $Bi_3Se_4Br$ (g,h). The positions of the conduction band minimum (CBM) and valence band maximum (VBM) are shown for all compounds.

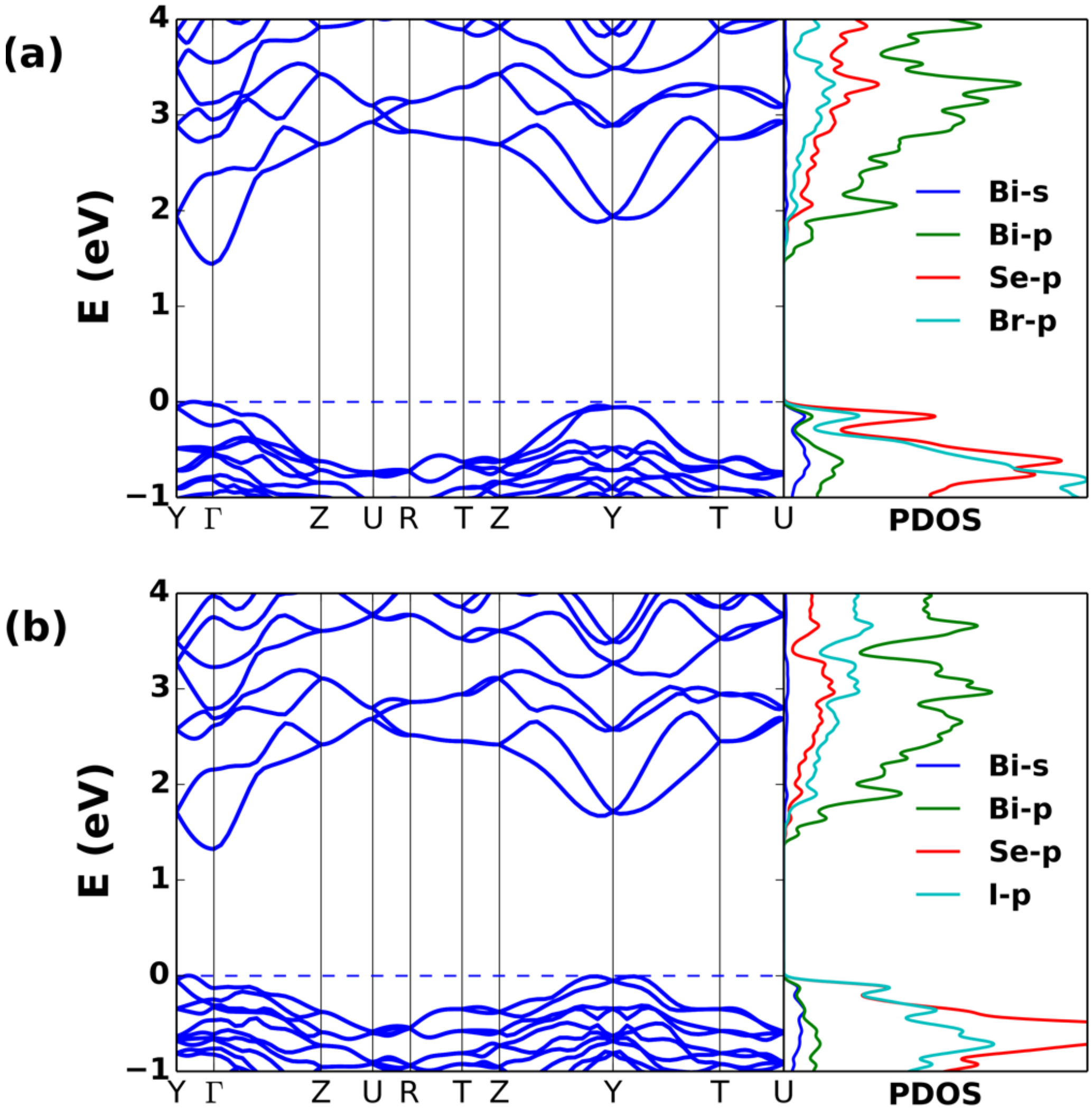

**Figure S2.** Band structures and partial electronic density of states of (a) BiSeBr and (b) BiSeI.

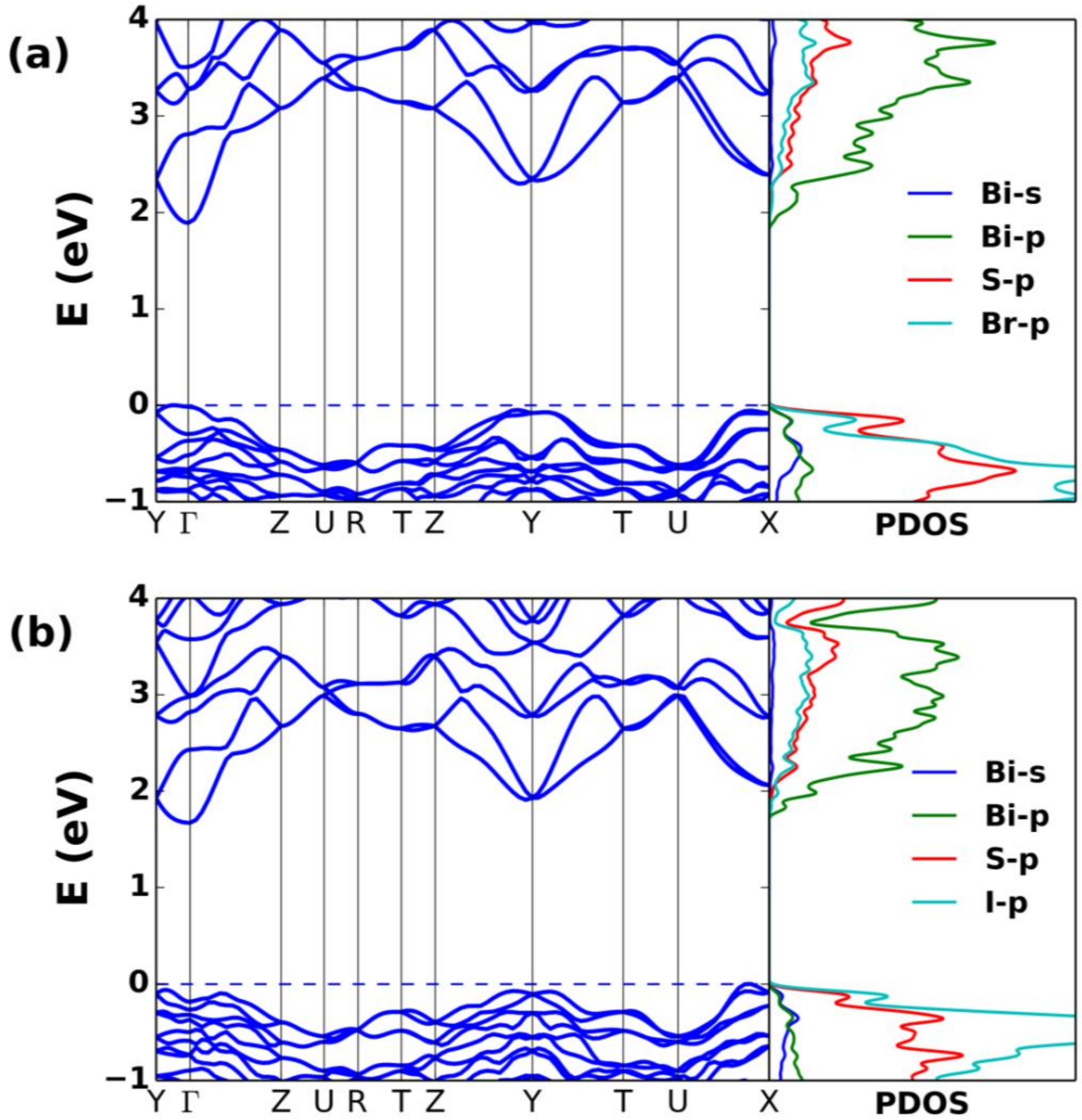

**Figure S3.** Band structures and partial electronic density of states of (a) BiSBr and (b) BiSI.

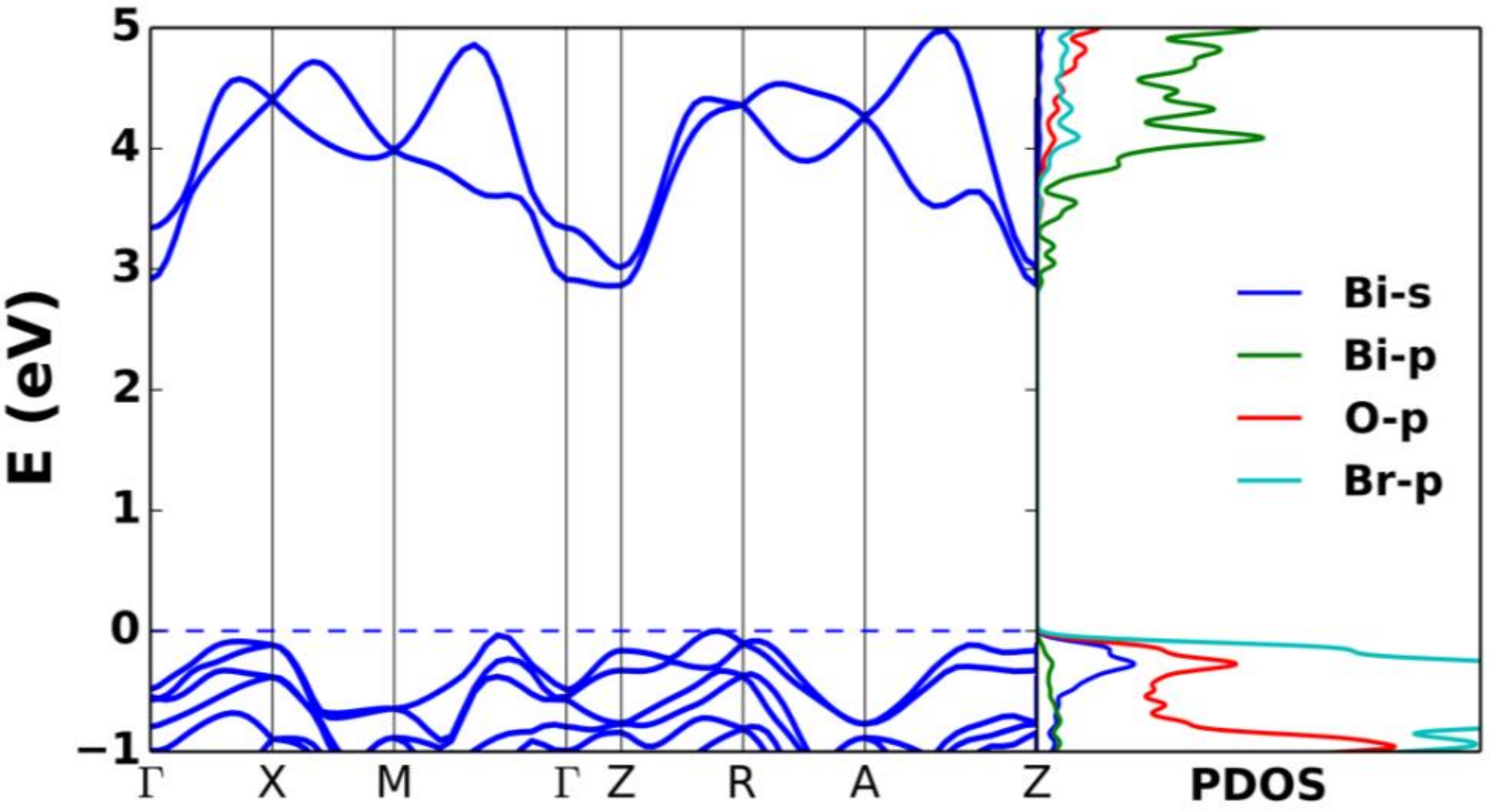

**Figure S4.** Band structures and partial electronic density of states of BiOBr.

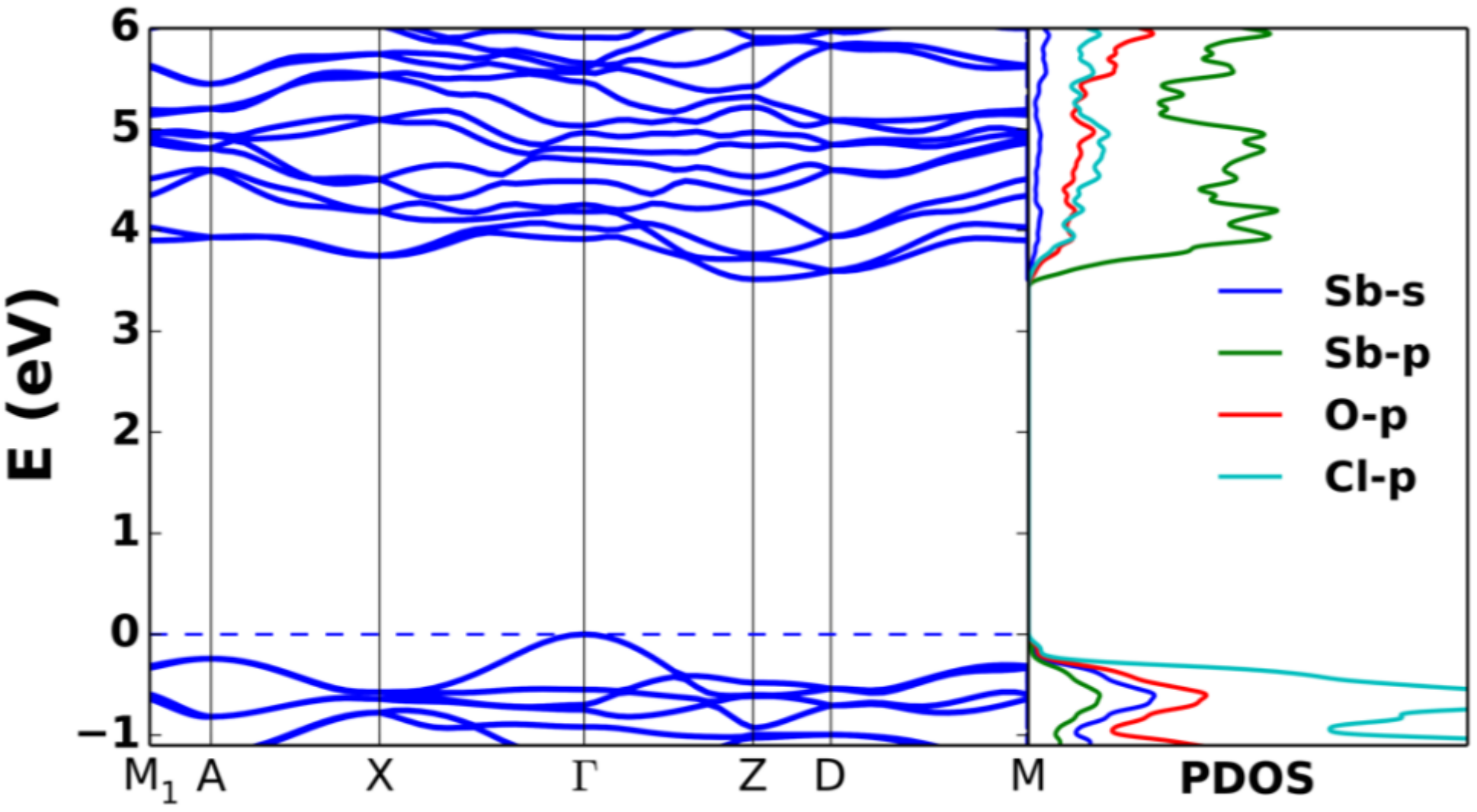

**Figure S5.** Band structures and partial electronic density of states of $Sb_4O_5Cl_2$.